\documentclass[fleqn,usenatbib]{mnras}

\usepackage{newtxtext,newtxmath}

\usepackage[T1]{fontenc}

\DeclareRobustCommand{\VAN}[3]{#2}
\let\VANthebibliography\thebibliography
\def\thebibliography{\DeclareRobustCommand{\VAN}[3]{##3}\VANthebibliography}

\usepackage{graphicx}	
\usepackage{amsmath}	
\usepackage{multirow}
\usepackage{xcolor}

\usepackage[print-unity-mantissa=false]{siunitx}
\DeclareSIUnit\angstrom{\textup{\AA}}
\DeclareSIUnit\erg{erg}
\DeclareSIUnit\parsec{pc}
\DeclareSIUnit\msol{M_{\odot}}
\DeclareSIUnit\year{yr}

\usepackage{physics}
\AtBeginDocument{\RenewCommandCopy\qty\SI}

\newcommand{\FeVII}{[Fe \textsc{vii}]}
\newcommand{\FeX}{[Fe \textsc{x}]}
\newcommand{\NeV}{[Ne \textsc{v}]}
\newcommand{\HeII}{He \textsc{ii}}

\newcommand{\cloudy}{\textsc{cloudy}}

\defcitealias{homan2023}{H23}

\title[Coronal Line Reverberation Mapping in MKN 110]{Long Term Reverberation Mapping of Iron Coronal Lines in MKN 110}

\author[C. Yin et. al.]{
Charles Yin,$^{1}$\thanks{E-mail: charlesyin42@gmail.com}
Andy Lawrence,$^{1}$
Martin Ward,$^{2}$
David Homan,$^{3,4}$
Wolfram Kollatschny$^{5}$
\\
$^{1}$Institute for Astronomy, SUPA (Scottish Universities Physics Alliance), University of Edinburgh, Royal Observatory, Blackford Hill, Edinburgh EH9 3HJ, UK\\
$^{2}$Department of Physics, Durham University, South Road, Durham DH1 3LE, UK\\
$^{3}$Leibniz-Institut f\"ur Astrophysik Potsdam, An der Sternwarte 16, 14482 Potsdam, Germany\\ 
$^{4}$Institute of Astronomy, University of Cambridge, Madingley Road, Cambridge CB3 0HA\\
$^{5}$Institut f\"ur Astrophysik und Geophysik, Universit\"at G\"ottingen, Friedrich-Hund Platz 1, 37077 G\"ottingen, Germany
}

\date{Accepted XXX. Received YYY; in original form ZZZ}

\pubyear{2024}

\begin{document}
\label{firstpage}

\pagerange{\pageref{firstpage}--\pageref{lastpage}}
\maketitle

\begin{abstract}
We present flux measurements of the coronal lines \FeVII{} and \FeX{} spanning three decades, in the highly variable Active Galactic Nucleus (AGN) MKN 110. These coronal lines are sensitive to the spectral energy distribution (SED) of AGNs in the extreme ultraviolet (EUV). Neither \FeVII{} nor \FeX{} demonstrates variability in the short term on a weekly or monthly timescale. However, by taking advantage of a long term decrease in the continuum flux of MKN 110 on the order of years, we were able to track the \FeVII{} and \FeX{} fluxes as they respond to the continuum. We were able to detect a lag for \FeVII{} relative to the continuum at \SI{5100}{\angstrom}, with a modal lag of 652 days, but were unable to detect a significant lag in the \FeX{} flux,  though there exist significant uncertainties in the \FeX{} fit. These two lag results are not consistent and the line widths for the two line species also do not match. This provides strong evidence for stratification within the coronal line region (CLR). There is also evidence of a non-varying component within the coronal line flux, probably a result of a more extended region of origin. Taken together, these results suggest a CLR where the bulk of the \FeVII{} originates on parsec scales, but a portion of the \FeVII\ flux originates further out, at or beyond a \SI{10}{\parsec} scale. These results also indicate the limitations of single-cloud models in describing the physical conditions of the CLR.
\end{abstract}




\section{Introduction} \label{sec:prev_work}

It has long been observed that the UV peak of AGNs, assumed to be associated with the accretion disk, both varies significantly faster and is at a significantly lower frequency than predicted by simple accretion disk models \citep{lawrence2012}. One appealing explanation is that the observed UV peak is a "false summit" caused by atomic emission, and that the bulk of the emission from the disk is actually at higher frequencies.
The presence of significant radiation in the extreme UV (\qtyrange{90}{400}{\electronvolt}) would produce highly ionized atoms, which can be studied by transitions from such highly ionized species which occur in the optical or near-infrared. These emission lines are commonly referred to as coronal lines, so called as they were first detected in the solar corona. These coronal lines are produced by forbidden and semi-forbidden atomic transitions \citep{appenzeller1991}, and therefore also referred to as Forbidden High Ionization Lines (FHILs).

Coronal lines have been detected in AGNs as far back as \citet{oke1968}, who first reported the presence of \FeX{} $\lambda$\SI{6375}{\angstrom} and [Fe \textsc{XIV}] $\lambda$\SI{5304}{\angstrom} in NGC 4151. Prominent coronal lines have been noted in observations of AGNs, with \citet{derobertis1984} finding coronal lines in 12 Seyferts including III Zw 77, and in the subject of this paper, MKN 110 by \citet{kollatschny2001a}. Although coronal lines are present in  AGN spectra, their presence is not universal. An early survey of coronal lines by \citet{penston1984} detected \FeX{} emissions in 12 out of 29 Seyfert galaxies. More recent surveys \citep{rodriguez-ardila2011,lamperti2017,muller-sanchez2018} found the fraction of AGNs with coronal line emission to be between $\sim\SI{30}{\percent}$ and $\sim\SI{90}{\percent}$, with fractions varying depending on the sample selection, the choice of coronal lines species, the signal to noise of the observation and the distance to the galaxies. Commonly observed coronal lines include [Si \textsc{vi}] $\lambda$\SI{1.963}{\micro\meter}, [S \textsc{viii}] $\lambda$\SI{0.991}{\micro\meter}, [Si \textsc{ix}] $\lambda$\SI{1.252}{\micro\meter} and [Si \textsc{x}] $\lambda$\SI{1.430}{\micro\meter} in the infrared, and \NeV{} $\lambda$\SI{3426}{\angstrom}, \FeVII{} $\lambda$\SI{6087}{\angstrom} and \FeX{} in the optical. However, only a few AGN display all the observed coronal lines simultaneously. This is typically attributed to the mass of the central black hole \citep[which should impact the shape of the SED - see e.g. ][]{reefe2022}, the effect of dust grains adsorbing iron, especially on iron lines \citep{negus2023,mckaig2024}, or obscuration \citep{gelbord2009} by a dusty "torus".

While it is quite common to detect coronal lines in AGNs, in normal galaxies coronal lines are rare \citep{reefe2022}. Since coronal lines are understood to originate via photoionization, they have been proposed as a method of searching for AGNs \citep{negus2023}, as the presence of coronal lines requires ionizing radiation above $\sim$\SI{100}{\electronvolt}, which is beyond that from hot stars and Type II Supernovae present in starburst galaxies. This is an alternative to identifying AGN based in broad permitted lines, which has limitations for classes such as Seyfert 2s and narrow line Seyfert 1s.

The CLR (Coronal Line Region) is thought to be stratified, with lines produced by species with higher ionization potentials being emitted closer to the central black hole. \citet{rodriguez-ardila2011} find that coronal line Full Width Half Maxima (FWHM) are larger in lines from species with higher ionization potentials, similar to what has been found in the BLR. \citet{mazzalay2010} finds that coronal line emissions from species with higher ionization potentials are more spatially concentrated. \citet{gelbord2009} also argue for a stratified CLR, based on the relative fluxes of \FeVII{} and \FeX{} in Seyfert 1 vs Seyfert 2, as in the latter type the more compact \FeX{} region is partially obscured. The correlation between the width and blueshift of coronal lines have also been used to infer the presence of outflows such as warm winds \citep{erkens1997}.

While coronal lines offer the potential to be a powerful diagnostic tool to study the EUV SED of AGNs, a problem encountered in modelling coronal lines is that there remain uncertainties about the radius at which coronal lines are emitted. Together with systematic studies and measurements of coronal lines in samples of AGNs there have also been spatially resolved observations of them, including of Circinus \citep{oliva1994}, a sample of 10 Seyferts by \citet{mazzalay2010} and 71 AGN by \citet{negus2023}. Such observations detected coronal line emission up to \SI{5}{\kilo\parsec} from the central source. In many cases, the bulk of the emission is concentrated in the central ($<$\SI{10}{\parsec}) region that is not spatially resolved. However, \citet{negus2023} found several objects where only extended emissions were found, and no nuclear spatial pixels show \FeVII{} emission, possibly due to dust grain effects. These off-nucleus emissions are often associated with powerful outflows. 

As modelling by among others \citet{ferguson1997} demonstrated, coronal lines can be emitted across a wide range of distances (see Section \ref{sec:modelling} for a discussion of this). If the bulk of coronal line emission arises at or within \SI{10}{\parsec} from the central source, it is plausible that we could observe coronal line fluxes to vary in response to the continuum, such as is found for the broad permitted emission lines. The variability of broad lines has been studied in great detail via reverberation mapping \citep[as described by][]{peterson2004}. In comparison variability of coronal lines has received much less attention. We shall discuss some examples in the following section.

\subsection{Variability of Coronal Lines}

One of the earliest systematic searches for coronal line variability was performed by \citet{veilleux1988}. Out of 18 objects with strong coronal lines, only one - NGC 5548 (subsequently studied in greater detail by \citet{landt2015} and \citet{kynoch2022}) was found to exhibit clear variability in both \FeVII{} and \FeX{}. Several other objects were claimed to exhibit possible evidence of coronal line variation, including NGC 4151, the subject of \citet{landt2015a}. \citet{veilleux1988} was limited in the search by the poor temporal sampling, with only 55 spectra available for their 18 objects. Most objects were observed over the duration of 3 to 5 years, and for several objects only 1 year of spectra was available.

As part of the study of the variability of MKN110, \citet{kollatschny2001a} noted that, in comparing the spectra obtained in that campaign with spectra obtained by \citet{bischoff1999}, both the \FeVII{} and \FeX{} lines had varied. However, they did not attempt to measure the flux of either line, citing the limited signal-to-noise.

\citet{landt2015a} collated and obtained spectra for NGC 4151 over eight years, measuring coronal lines both in the optical, near-infrared and X-ray. They found that coronal lines in general only varied weakly. Towards the later part of the study period, it was noted that lines with the highest ionization potential weakly followed the broad line variations.

\citet{landt2015} carried out a comparable study of NGC 5548. Similar to \citet{landt2015a}, no strong evidence of coronal line variability was found. Indeed, while hot dust and infrared broad lines increased over a period of two years, no response was observed in the coronal line fluxes. However, this increase occurred towards the end of the observational period, so may not be fully reflected in the coronal line fluxes.

More recently, \citet{kynoch2022} carried out a 1-year IR monitoring campaign on NGC 5548, which found highly variable wings in the infrared coronal line profiles of [S VIII] and [Si VI], but otherwise, little variation in the core of the line. This suggests the presence of two distinct line emitting regions 

The other recent development was the discovery of coronal lines in TDEs AT2017gge \citep{onori2022} and AT2019qiz \citep{short2023}. These studies demonstrated that coronal lines can appear in response to Tidal Disruption Events. For instance, in AT2019qiz, \FeVII{} and \FeX{} emission only began to appear around 400 days after the outburst. \citet{short2023} was notable for having the benefit of a relatively well-sampled series of spectral observations, and so the authors were therefore able to observe the coronal lines responding to the outburst, as had been proposed by \citet{komossa2008}. Based on such evidence of \FeVII{} and \FeX{} responding to continuum changes associated with the central black hole, a sufficiently long time series of spectral observations could be used to study  the coronal line response in AGNs.

\subsection{MKN 110 - A Highly Variable AGN}

Most AGN display some variation in their optical spectrum across multiple timescales, but MKN 110 is particularly variable, displaying significant variation on a weekly timescale \citep[e.g.][ in addition to the papers listed in Table \ref{tab:spec}]{kollatschny2002}, daily \citep[e.g.][]{vincentelli2021} and possibly even on an hourly timescale \citep{webb2000}. Variability is also present into the X-Ray, UV and infrared \citep{vincentelli2021,vincentelli2022}.

MKN 110 has a redshift of $z=0.0353$, corresponding to \SI{150}{\mega\parsec}. \citet{meyer-hofmeister2011} derive a bolometric luminosity 0.4 times the Eddington limit, but this is based on a $\log(M_{BH}/M_{\odot})$ of 7.40, or \SI{2.51e7}{\msol}. This value was derived from reverberation mapping data by \citet{peterson2004}, and is supported by other reverberation mapping campaigns such as by \citet{kollatschny2001a} (\SI{1.83e7}{\msol}) and measurements based on the galactic bulge velocity dispersion such as \citet{onken2004}, which gives \SI{2.5e7}{\msol}. However, these figures are in effect a lower bound on the mass of the central black hole due to inclination effects. \citet{kollatschny2003} interpreted the observed redshift of broad emission lines as being the result of gravitational redshift, yielding a mass of \SI{1.4e8}{\msol}. This much higher result is in agreement with the spectropolarimetry based measurements by \citet{afanasiev2019} of \SI{2.1e8}{\msol}. The discrepancy between the gravitational redshift mass \citep{kollatschny2003} and reverberation mapping mass \citet{kollatschny2001a} implies an inclination of \SI{21}{\deg} from pole-on.

The optical continuum at \SI{5100}{\angstrom} of MKN 110 has dimmed consistently and appreciably by a factor of $\sim3$ between MJD 52000 and MJD 54000. In this paper, we use the large set of spectra now available to search for a response in the coronal line fluxes related to this dip in continuum flux. This extends the study of \citet{homan2023} (hereafter referred to as \citetalias{homan2023}), into the properties of the coronal line region.

This paper is organized thus: Section \ref{sec:dataset} describes the set of archival spectra used in this paper. Section \ref{sec:method} outlines the process of fitting the coronal lines, including the \FeVII{} and \FeX{} lines. Section \ref{sec:width} compares the line profile of \FeVII{} with other emission lines. Sections \ref{sec:variability} and \ref{sec:lag} demonstrate that the \FeVII{} flux varies in response to the continuum, and quantifies the time lag in the response. Section \ref{sec:responsivity} examines the responsivity of the \FeVII{} line. Finally, Section \ref{sec:discussion} discusses these results in the context of understanding the physical conditions within the CLR.

\section{Datasets Used} \label{sec:dataset}

In this paper we use the same data set as \citetalias{homan2023}, which we briefly summarize below. While \citetalias{homan2023} were primarily concerned with the line flux and velocity measurements of some strong optical emission lines, the original spectral data were available only for a portion of the observations, including SDSS and newer. For this paper we also use the original reduced spectra analysed by \citet{kollatschny2001a}.

The spectra in \citetalias{homan2023} were assembled from multiple previous papers, and the analysis in \citetalias{homan2023} was based on both previously published line fluxes and new measurements reported in that paper. In this paper we present new measurements of the iron coronal lines which have not previously been reported. For consistency with \citetalias{homan2023}, we adopt the same shorthand notation to refer to each data set. The list of spectra and the corresponding references are listed in Table \ref{tab:spec}. Further details on each data set can be found in \citetalias{homan2023}.

\begin{table*}
	\centering
	\caption{A summary of the various sources of spectra used in this paper.}
	\label{tab:spec}
	\begin{tabular}{llcclll}
		\hline
		                         &            & \multicolumn{2}{c}{Date Range (MJD)} & No. of  & \multicolumn{2}{c}{Wavelength Range (\unit{\angstrom})} \\
		Source                   & Shorthand  & First             & Last             & spectra & Blue/Min          & Red/Max                             \\
		\hline
		\citet{bischoff1999}     & BK99       & 46828             & 49870            & 13      & 4249              & 7049                                \\
		                         &            & (Feb 1987)        & (Jun 1995)       &         &                   &                                     \\
		\citet{kollatschny2001a} & K01        & 51495             & 51678            & 26      & 4125.52           & 7175.02                             \\
		                         &            & (Nov 1999)        & (May 2000)       &         &                   &                                     \\
		                         & SDSS       & \multicolumn{2}{c}{52252}            & 1       & 3673.11           & 8900.90                             \\
		                         &            & \multicolumn{2}{c}{(Dec 2001)}       &         &                   &                                     \\
		                         & FAST-RM1   & 52581             & 52798            & 26      & 3535.88           & 7272.93                             \\
		                         &            & (Nov 2002)        & (Jun 2003)       &         &                   &                                     \\
		                         & FAST-RM2   & 52940             & 53137            & 22      & 3538.97           & 7278.00                             \\
		                         &            & (Oct 2003)        & (May 2004)       &         &                   &                                     \\
		\citet{landt2008}        & FAST-Landt & 53741             & 54116            & 2       & 3359.66           & 7158.27                             \\
		\citet{landt2011}        &            & (Jan 2006)        & (Jan 2007)       &         &                   &                                     \\
		                         & FAST-extra & 54122             & 55179            & 2       & 3357.01           & 7158.36                             \\
		                         &            & (Jan 2007)        & (Dec 2009)       &         &                   &                                     \\
		                         & M15        & 56450             & 57369            & 3       & 3255.19           & 7696.27                             \\
		                         &            & (Jun 2013)        & (Dec 2015)       &         &                   &                                     \\
		                         & WHT        & 57539             & 57801            & 2       & 2994.36           & 10238.38                            \\
		                         &            & (May 2016)        & (Feb 2017)       &         &                   &                                     \\
		                         & K17        & \multicolumn{2}{c}{57834}            & 1       & 4476.96           & 6713.03                             \\
		                         &            & \multicolumn{2}{c}{(Mar 2017)}       &         &                   &                                     \\
		                         & FAST-new   & 58281             & 58555            & 9       & 3362.02           & 7168.32                             \\
		                         &            & (Jun 2018)        & (Mar 2019)       &         &                   &                                     \\
		\hline
	\end{tabular}
\end{table*}

\subsection{BK99}

This set of spectra from \citet{bischoff1999} contain observations taken between February 1987 and June 1995, made at the Calar Alto Observatory in Spain using the 2.2 m and 3.5 m telescopes, in addition to those from the McDonald Observatory in Texas, using the 2.1 and 2.7 m telescopes. \citet{bischoff1999} report a wavelength pixel scale of between 3 and \SI{7}{\angstrom}. This range of scales is due to the use of different telescopes and sensors.

\subsection{K01}

This set of spectra from \citet{kollatschny2001a} was obtained using the 9.2m Hobby-Eberly Telescope (HET) and the Marcario Low Resolution Spectrograph (LRS) at the McDonald Observatory between November 1999 and May 2000. The spectral resolution is 650 at \SI{4000}{\angstrom}.

\subsection{SDSS}

This spectrum was taken in 2001, and published as part of the Sloan Digital Sky Survey (SDSS) DR16. This spectrum is particularly important as it was obtained when the optical continuum and \HeII{} $\lambda$\SI{4686}{\angstrom} flux (which \citetalias{homan2023} argued was a better tracer of the UV continuum) from MKN 110 were at a minimum. The spectral resolution is 1500 at \SI{3800}{\angstrom}.

\subsection{FAST}

These sets of spectra were obtained using the FAST spectrograph on the Tillinghast 60-inch telescope at Mount Hopkins in Arizona, some of which are reported by \citet{landt2008} and \citet{landt2011}. The spectral resolution is 1600 at \SI{4800}{\angstrom}.

\subsection{M15}

These spectra were obtained using the 122cm Asiago telescope in 2013 and 2015 by Marco Berton.

\subsection{WHT}

This data set consists of two spectra taken by \citetalias{homan2023} in May 2016 and February 2017 using the ISIS spectrograph on the 4.2m William Hershel Telescope. The spectral resolution is at \SI{5200}{\angstrom}.

\subsection{K17}

This spectrum was obtained using the HET and LRS2-B spectrograph in 2017. The spectral resolution is 1140 for the orange arm, which covers \qtyrange{4600}{7000}{\angstrom}.

\section{Line Fitting} \label{sec:method}

For each spectrum a pseudo-continuum was fitted around the lines and subtracted. This pseudo-continuum accounts for both the shape of the overall continuum, extended wings from nearby strong broad emission lines (for instance the \FeVII{} sits on the extended blue wings of the broad He I $\lambda$\SI{5876}{\angstrom} line), as well as the presence of numerous blended fainter emission lines and features which cannot be confidently identified.

The pseudo-continuum is calculated as follows: we took a section of the continuum on the red-ward and blue-ward side of the emission line, divided it into four sections, averaged the flux, and interpolated with a cubic spline. For \FeVII{} we used spectra from \qtyrange{5900}{6048}{\angstrom} and from \qtyrange{6105}{6250}{\angstrom}. For \FeX{} we used spectra from \qtyrange{6325}{6350}{\angstrom} and from \qtyrange{6394}{6425}{\angstrom}. This produced a model of the underlying continuum (along with all features not associated with the emission line itself) with minor features smoothed out, allowing us to isolate the emission feature.

Some spectra, especially the FAST spectra, were too noisy to obtain a good fit for \FeVII{}. To overcome this we averaged several spectra together to produce a stacked spectra. In the case of FAST-RM1 and -RM2, we divided the spectra into three groups each - from MJD 52581 to 52648, from MJD 52651 to 52704, from MJD 52722 to 52798, from MJD 52940 to 53005, from MJD 53016 to 53076 and from MJD 53082 to 53137. For FAST-new, we combined all spectra into one stacked spectrum. We claim that this technique is justified since we do not expect the coronal lines to respond to changes in the continuum on the timescale of weeks, and indeed later show evidence confirming this assumption. The FAST-RM1 stacks contain spectra taken over the course of $\sim$60 days. The FAST-RM2 stacks contain spectra taken over the course of $\sim$40 days. The FAST-new stacked spectrum contains spectra taken over the course of 125 days.

For fitting, we use the package \texttt{lmfit}.

\subsection{Fitting of the \FeVII{} Line}

A two-Gaussian model is required to take account of the presence of a telluric feature which is close enough to the \FeVII{} to likely affect attempts to calculate the line flux by direct integration.

\begin{figure}
	\includegraphics[width=\columnwidth]{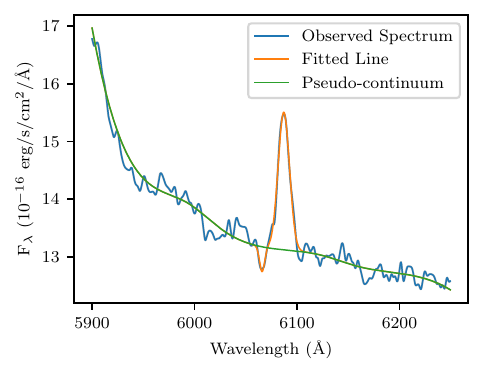}
    \caption{The pseudo-continuum as well as the \FeVII{} line fit shown with the observed spectrum for a sample spectrum (taken on MJD 51605)}
    \label{fig:line_fit}
\end{figure}

\begin{figure}
	\includegraphics[width=\columnwidth]{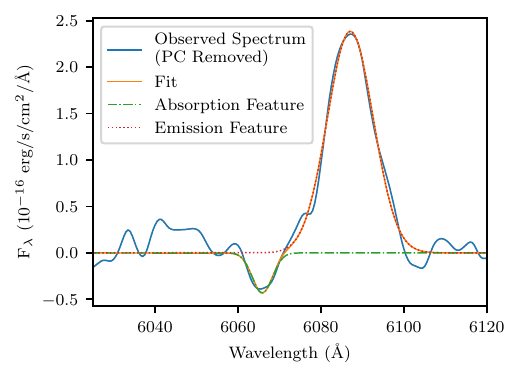}
    \caption{Demonstrating the process used for fitting both the telluric feature as well as the \FeVII{} emission in a sample spectrum (taken on MJD 51605). Here the pseudo-continuum has been subtracted from the observed spectrum.}
    \label{fig:fit_components}
\end{figure}

An example of our \FeVII{} line fit, including the line, the absorption feature and the pseudo-continuum is shown in Figure \ref{fig:line_fit}. The individual components including the absorption feature is shown in Figure \ref{fig:fit_components}, where the pseudo-continuum has been removed. The line fluxes quoted below represent the integral of the positive Gaussian component (the emission feature fitted by the red dashed Gaussian in Figure \ref{fig:fit_components}).

\subsection{Fitting of the \FeX{} Line}

The \FeX{} line is a relatively weak emission feature that is blended with the often (but not always) stronger [O \textsc{i}] $\lambda$\SI{6363}{\angstrom} emission line. Fitting a free two-Gaussian model, with one Gaussian for each component, is problematic as this results in a strong correlation between the [O \textsc{i}] $\lambda$\SI{6363}{\angstrom} line flux and the \FeX{} line flux (a correlation $\sim-0.96$). Also, the \FeX{} line is not necessarily the strongest contribution to the wings of the [O \textsc{i}] $\lambda$\SI{6363}{\angstrom} line. Fortunately, the [O \textsc{i}] $\lambda$\SI{6363}{\angstrom} line is known to be exactly 0.323 times the flux of the prominent [O \textsc{i}] $\lambda$\SI{6300}{\angstrom} line, which is free from blending from blending of other features of significant strengths. By fixing the ratio of [O \textsc{i}] $\lambda$\SI{6300}{\angstrom} and [O \textsc{i}] $\lambda$\SI{6363}{\angstrom} fluxes, and subtracting the latter from the total flux of the blend,  we can obtain the strength of the \FeX{} line.

\begin{figure}
	\includegraphics[width=\columnwidth]{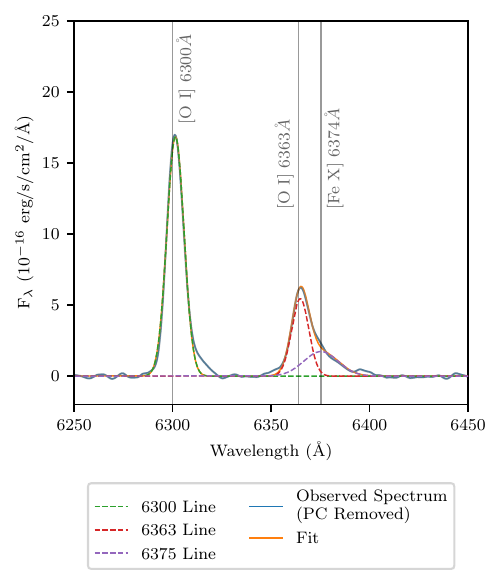}
    \caption{Demonstrating the process for fitting the \FeX{} line. The [O \textsc{i}] $\lambda$\SI{6300}{\angstrom}, [O \textsc{i}] $\lambda$\SI{6363}{\angstrom} and \FeX{} profiles are shown against the observed spectrum for a sample spectrum (MJD 51605).}
    \label{fig:fe10_fit_components}
\end{figure}

\begin{figure}
	\includegraphics[width=\columnwidth]{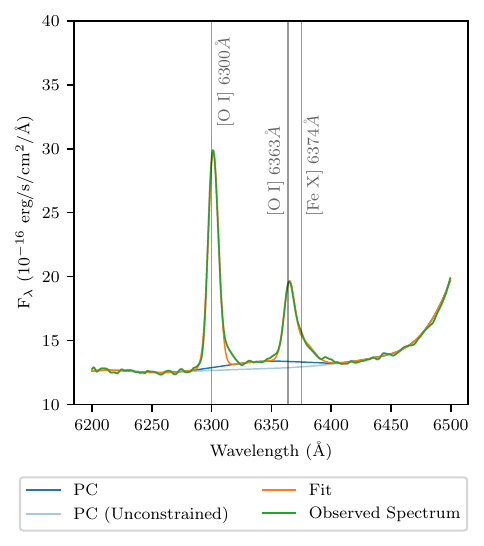}
    \caption{The pseudo-continuum as well as the [O \textsc{i}] $\lambda$\SI{6300}{\angstrom}, [O \textsc{i}] $\lambda$\SI{6363}{\angstrom} and \FeX{} line fit compared against the observed spectrum for a sample spectrum (taken on MJD 51605). The pseudo-continuum shown includes the strongly rising blue wing of the H$\alpha$ line.}
    \label{fig:fe10_line_fit}
\end{figure}

Our \FeX{} line fit, including the [O \textsc{i}] $\lambda$\SI{6300}{\angstrom}, [O \textsc{i}] $\lambda$\SI{6363}{\angstrom}, \FeX{} line components and the pseudo-continuum are shown in Figure \ref{fig:fe10_line_fit}. 
The individual emission lines which make up this blended complex are shown in Figure \ref{fig:fe10_fit_components} with the pseudo-continuum removed.

In the process described above, the portion of the spectrum between \SI{6322}{\angstrom} and \SI{6348}{\angstrom} is assumed to be part of the pseudo-continuum as opposed to the extended wings of either [O \textsc{i}] $\lambda$\SI{6300}{\angstrom} or [O \textsc{i}] $\lambda$\SI{6363}{\angstrom}. We repeated the constrained fitting of the two [O \textsc{i}] and \FeX{} lines but without assuming that the spectrum between \SI{6322}{\angstrom} and \SI{6348}{\angstrom} (between the two [O \textsc{i}] lines) is unrelated to the two [O \textsc{i}] lines, excluding it from our calculations for the pseudo-continuum. We found that this made a modest difference (consistently about 25\%) to the measured fluxes. The two different methods of fitting the pseudo-continuum are compared in Figure \ref{fig:fe10_line_fit}.

\begin{figure}
	\includegraphics[width=\columnwidth]{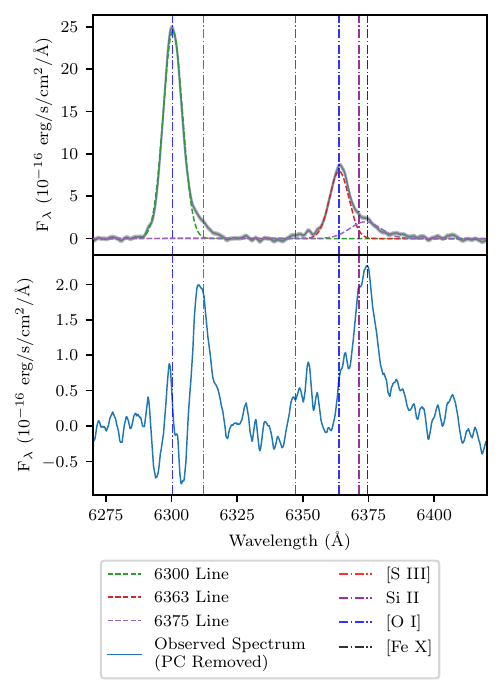}
    \caption{Demonstrating the confirmation of \FeX{} emission, here using a stacked spectrum combined both FAST-RM1 and FAST-RM2 spectra. The top panel shows the fit, with the pseudo-continuum removed. The bottom panel shows the spectrum with the [O \textsc{i}] components subtracted. We highlight the expected positions of key lines, colour-coded by species. The wavelengths marked correspond to [O \textsc{i}] $\lambda$\SI{6300}{\angstrom}, [O \textsc{i}] $\lambda$\SI{6363}{\angstrom}, [s \textsc{iii}] $\lambda$\SI{6313}{\angstrom}, Si \textsc{ii} $\lambda$\SI{6347}{\angstrom}, Si \textsc{ii} $\lambda$\SI{6371}{\angstrom} and \FeX{}.}
    \label{fig:o6300_residue}
\end{figure}

There have been cases of lines being misidentified as \FeX{} instead of as Si \textsc{ii} $\lambda$\SI{6371}{\angstrom}, for instance as described by \citet{herenz2023}. We check for this explicitly as demonstrated in figure \ref{fig:o6300_residue}. We remove the [O \textsc{i}] emissions, and despite some remaining artefacts around where we have removed [O \textsc{i}] $\lambda$\SI{6300}{\angstrom} attributable to an imperfect fit, we are able to identify a [S \textsc{iii}] $\lambda$\SI{6313}{\angstrom} component. We also find that any Si \textsc{ii} $\lambda$\SI{6347}{\angstrom} emission is significantly weaker than that of the likely \FeX{} emission. Given that the ratio of Si \textsc{ii} $\lambda$\SI{6347}{\angstrom} to Si \textsc{ii} $\lambda$\SI{6371}{\angstrom} appears to be approximately 1:1 \citep{herenz2023} it is likely that Si \textsc{ii} $\lambda$\SI{6371}{\angstrom} is a small contribution to the line. We therefore conclude that the other main emission is more consistent with being \FeX{} rather than Si \textsc{ii} $\lambda$\SI{6371}{\angstrom}. Given that we find \FeX{} varies by a factor of up to $\sim$5, variability in \FeX{} is very likely not a result of Si \textsc{ii} $\lambda$\SI{6371}{\angstrom} variability.

\subsection{Other Coronal Lines}

\begin{figure}
	\includegraphics[width=\columnwidth]{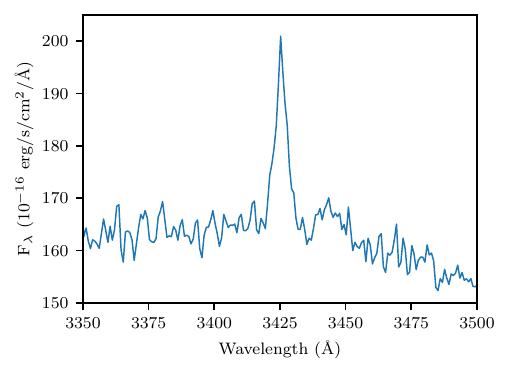}
    \caption{An example of the \NeV{} emission line in a sample spectrum where the blue optical is available (taken on MJD 57539).}
    \label{fig:ne5_line_fit}
\end{figure}

The FAST-Landt, FAST-extra, WHT and FAST-new spectra extended into the far blue optical, where we can detect the \NeV{} emission line. Unfortunately, given the limited number of observations that cover this line, it is not possible to conclude whether the \NeV{} line exhibits variability similar to that found in the \FeVII{} line. An example of the \NeV{} emission line is shown in Figure \ref{fig:ne5_line_fit}.

\begin{figure}
	\includegraphics[width=\columnwidth]{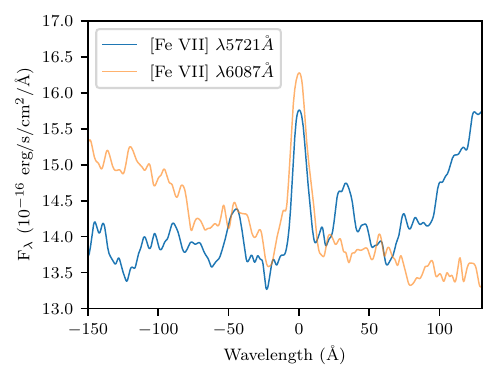}
    \caption{The [Fe VII] $\lambda$\SI{5721}{\angstrom} line in a sample spectrum (taken on MJD 51605), compared with the \FeVII{} line. The \FeVII{} line is shifted vertically, but not scaled.}
    \label{fig:fe7_5721_line_fit}
\end{figure}

We were also able to find evidence for the presence of the [Fe \textsc{vii}] $\lambda$\SI{5721}{\angstrom} line. This line was less distinct but it has the same atomic upper level state as the \FeVII{} line, and so the ratio of the flux of the [Fe \textsc{vii}] $\lambda$\SI{5721}{\angstrom} line to \FeVII{} is fixed at 0.61. The profiles of the two lines are compared in Figure \ref{fig:fe7_5721_line_fit}.

We were able to identify several other forbidden high ionization lines, including [Ne \textsc{iii}] $\lambda$\SI{3969}{\angstrom}, [Ca \textsc{v}] $\lambda$\SI{5309}{\angstrom} and [Ar \textsc{iii}] $\lambda$\SI{7136}{\angstrom}. However, these lines are generally quite weak, with numerous other nearby emission and/or absorption features, limiting our confidence in any fit. Given the complexity of the spectra around these lines, as well as the relatively low ionzation energy of these species, we be focusing our discussion on \FeVII and \FeX emissions.

\section{The Width and Profile of the Coronal Lines} \label{sec:width}

\begin{table}
	\centering
	\caption{Line Widths and corresponding Doppler velocity of \FeVII{} in various spectra corrected for instrumental broadening.}
	\label{tab:fe7_widths}
	\begin{tabular}{lcccc}
		\hline
		              &      & FWHM       & Velocity    & Velocity                      \\
		Shorthand     & R    & (\AA)      & Width (\AA) & (\si{\kilo\meter\per\second}) \\
		\hline
		K01           & 650  & 11.45      & 6.34          & 312                           \\
		SDSS          & 1500 &  8.64      & 7.63          & 376                           \\
		FAST-RM1 \& 2 & 1600 &  8.27      & 7.34          & 361                           \\
		WHT           & 1500 &  8.91      & 7.87          & 387                           \\
		FAST-New      & 1600 &  9.34      & 8.53          & 420                           \\
		\hline
		Mean          &      &            & 6.69          & 330                           \\
		\hline
	\end{tabular}
\end{table}

\begin{table}
	\centering
	\caption{Line Widths and corresponding Doppler velocity of \FeX{} in various spectra corrected for instrumental broadening.}
	\label{tab:fe10_widths}
	\begin{tabular}{lcccc}
		\hline
		              &      & FWHM       & Velocity    & Velocity                      \\
		Line          & R    & (\AA)      & Width (\AA) & (\si{\kilo\meter\per\second}) \\
		\hline
		K01           & 650  & 19.43      & 16.70       & 786                           \\
		SDSS          & 1500 & 13.17      & 12.47       & 586                           \\
		FAST-RM1 \& 2 & 1600 & 15.19      & 14.48       & 681                           \\
		WHT           & 1500 & 21.16      & 20.72       & 975                           \\
		FAST-New      & 1600 & 21.19      & 20.82       & 979                           \\
		\hline
		Mean          &      &            & 16.55       & 778                           \\
		\hline
	\end{tabular}
\end{table}

\begin{table}
	\centering
	\caption{Line Widths and corresponding Doppler velocity of [O \textsc{i}] 6300 and [O \textsc{iii}] 5007 in various spectra corrected for instrumental broadening.}
	\label{tab:o_widths}
	\begin{tabular}{lcccc}
		\hline
		              & \multicolumn{2}{c}{[O \textsc{i}] 6300}               & \multicolumn{2}{c}{[O \textsc{iii}] 5007}             \\
		              & Observed   & Velocity                      & Observed   & Velocity                      \\
		Line          & FWHM (\AA) & (\si{\kilo\meter\per\second}) & FWHM (\AA) & (\si{\kilo\meter\per\second}) \\
		\hline
		K01           & 10.35      & 173                           & 9.20       & 301                           \\
		SDSS          &  7.62      & 303                           & 5.61       & 270                           \\
		FAST-RM1 \& 2 &  8.56      & 362                           & 7.51       & 409                           \\
		WHT           &  9.80      & 421                           & 5.98       & 297                           \\
		FAST-New      &  7.75      & 318                           & 7.56       & 412                           \\
		\hline
		Mean          &            & 224                           &            & 321                           \\
		\hline
	\end{tabular}
\end{table}

\begin{figure}
	\includegraphics[width=\columnwidth]{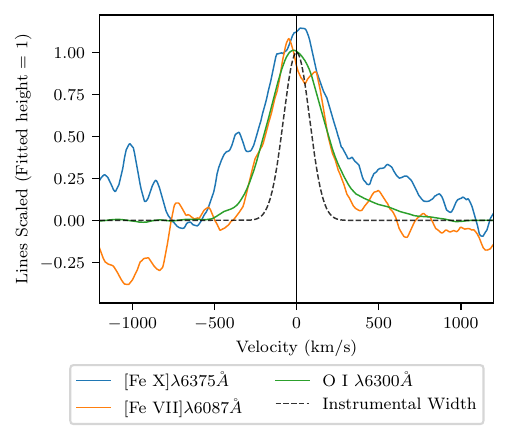}
    \caption{The [O \textsc{i}] $\lambda$\SI{6300}{\angstrom}, \FeVII{} and \FeX{} lines from the FAST-RM1 and FAST-RM2 spectra. The \FeX{} profile is generated by removing the fitted [O \textsc{i}] $\lambda$\SI{6363}{\angstrom} component. The flux is normalized such that the height of the fitted line is 1. The instrumental velocity width is shown for comparison.}
    \label{fig:line_overplot}
\end{figure}

While the line velocity widths of both \FeVII{} and \FeX{} are wider than the expected instrumental velocity width, from the available spectra we are unable to resolve features in the \FeVII{} and \FeX{} lines. We tabulate the widths of \FeVII{} in Table \ref{tab:fe7_widths}, the widths of \FeX{} in Table \ref{tab:fe10_widths}, and the widths of [O \textsc{i}] $\lambda$\SI{6300}{\angstrom} and [O \textsc{iii}] $\lambda$\SI{5007}{\angstrom} in Table \ref{tab:o_widths}. An example of the profiles of \FeVII{}, \FeX{} and [O \textsc{i}] $\lambda$\SI{6300}{\angstrom} are shown in Figure \ref{fig:line_overplot}, along with the instrumental width shown for comparison. It is clear the width of \FeX{} is greater than that of \FeVII{}. In principle it would be interesting to see if the profiles change from epoch to epoch. However, in light of the limited spectral resolution, our discussion of variability will be limited to the fluxes of \FeVII{} and \FeX{}.

\section{Variability of the Coronal Lines} \label{sec:variability}

\subsection{Variability of the \FeVII{} Line}

\begin{figure}
	\includegraphics[width=\columnwidth]{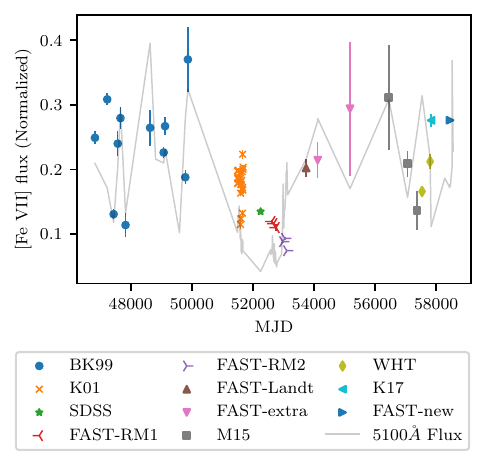}
    \caption{Fitted fluxes for the \FeVII{} emission lines, and the errors from fitting. The F5100 flux measured in \citetalias{homan2023} is shown for comparison.  All fluxes are normalized against the narrow [O \textsc{i}] $\lambda$\SI{6300}{\angstrom} line flux. However the fitting error does not cover the overall variance due to differences in resolution. Each separate observing run is colour coded and labelled.}
    \label{fig:time_series}
\end{figure}

Repeating the process described above on every spectrum we have available, including the stacked spectra as appropriate, we find that the \FeVII{} flux displays strong variability. The emission line flux and its associated error from the fitting process is shown in Figure \ref{fig:time_series}. A small number of spectra have very large errors ($\sigma_{F_{\lambda}} > 0.4F_{\lambda}$), and so are excluded. These spectra have unusually broad but weak \FeVII{} lines. In one case, the \FeVII{} line is completely absent. These problematic spectra are also significantly noisier than other spectra included in the same observational series. These large \FeVII{} errors are therefore likely a result of observational effects such as atmospheric conditions.

The error bars shown in Figure \ref{fig:time_series} do not capture all sources of the possible error in measuring the \FeVII{} flux. The fitting process is sensitive to the pseudo-continuum fitting, and the BK99 spectra in particular are of relatively low spectral resolution. Since the spectra used were obtained from a wide range of instruments over the course of 20 years, there are other systematic errors which are difficult to fully quantify, including those arising from flux calibrations, as well as aperture effects. The latter is especially important as extended coronal line emission, including that of \FeVII{} is common in AGN. We mitigate against the effect of these sources of uncertainty by normalising against the narrow [O \textsc{i}] $\lambda$\SI{6300}{\angstrom} line in Figure \ref{fig:time_series}, dividing by the [O \textsc{i}] $\lambda$\SI{6300}{\angstrom} flux from the same spectrum.

Visual inspection suggests that between MJD 52000 (April 2001) and MJD 53500 (May 2005), the \FeVII{} flux was appreciably lower than both before and afterwards, showing a factor $\sim$ 3 variation in normalized flux. This compares to the roughly 10\% difference in the fluxes of narrow lines such as [O \textsc{iii}] $\lambda$\SI{5007}{\angstrom}, which we consider to be an estimate of the scale of the systematic errors.

\begin{figure}
	\includegraphics[width=\columnwidth]{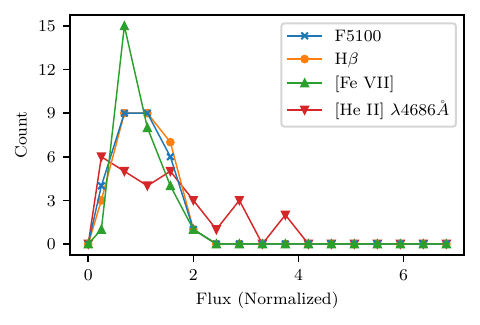}
    \caption{A comparison of the distribution of \FeVII{}, F5100, \HeII{} and H$\beta$ fluxes, all normalized to their respective values at MJD 47574. As K01 spans a short period of time, for this figure only, we have stacked the K01 data set to avoid the large number of similar fluxes distorting the histogram. The width of the distribution of normalized fluxes reflects the amount of variability displayed by these lines. The \FeVII distribution is notably narrower than the distribution for other lines, demonstrating that \FeVII displays much smaller variability than classic broad lines or the continuum.}
    \label{fig:hist}
\end{figure}

In Figure \ref{fig:hist} we show the measured line fluxes for \FeVII{} compared with the \SI{5100}{\angstrom} continuum (F5100, see \citetalias{homan2023} for more details), along with the \HeII{} and H$\beta$ line fluxes as a histogram. The range of variation in \FeVII{}, manifesting as the width of the distribution, is noticeably smaller than that of the broad lines measured by \citetalias{homan2023}. The F5100 and \HeII{} show greater variation in flux, and so the distribution of fluxes in Figure \ref{fig:hist} is wider. \citetalias{homan2023} found that \HeII{} has the shortest response time and the greatest variation in flux, and this manifests as  the \HeII{} distribution in Figure \ref{fig:hist} being the widest.

\begin{figure}
	\includegraphics[width=\columnwidth]{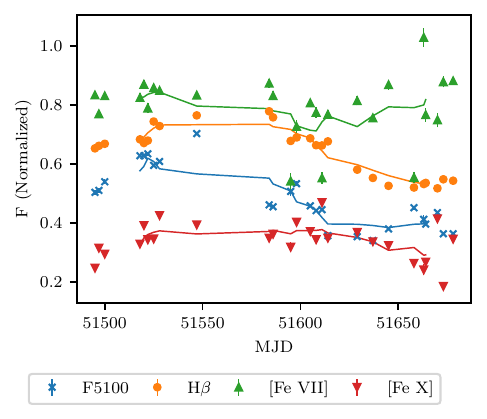}
    \caption{Short term variability of the F5100 continuum, H$\beta$, \FeVII{} and \FeX{} line fluxes from the K01 set of spectra. All fluxes are normalized against their respective values at MJD 47574 in keeping with \citetalias{homan2023}. The lines are 7-epoch moving averages, included to demonstrate general trends without the epoch-to-epoch scatter. The H$\beta$ flux declines over the duration of this data set in response to the F5100 flux, but the \FeVII{} flux does not change appreciably, as a result of the much longer \FeVII{} response timescale. The last 4 or 5 \FeX{} measurements appear to be lower in the latter half of the K01 observational time period, possibly related to the declining F5100 flux.}
    \label{fig:K01_time_series}
\end{figure}

The K01 series of spectra, spanning 183 days, provides an excellent opportunity to search for short term variations in the \FeVII{} flux, especially as all of the spectra are of high quality. As previously discussed in greater detail in \citet{kollatschny2001a} and later utilized in \citetalias{homan2023}, the continuum and broad line fluxes do significantly decrease over time, and the H$\beta$ fluxes tracks the continuum. However, this is not the case for the \FeVII{} flux, as shown in Figure \ref{fig:K01_time_series}.

More rigorously, taking the number of flux measurements among the first 13 epochs that lie above average for K01, and the number among the second 13 epochs that lie below average, we compute the probability that this arises purely by chance. For F5100 and H$\beta$, these are \SI{1.37e-6} and \SI{1.63e-5}, while for \FeVII{} this is 0.0667, and for \FeX{} 0.206.

In summary, although the \FeVII{} flux varies on longer timescales, there is little evidence for variations on shorter timescales.

\subsection{Variability of the \FeX{} Line}

\begin{figure}
	\includegraphics[width=\columnwidth]{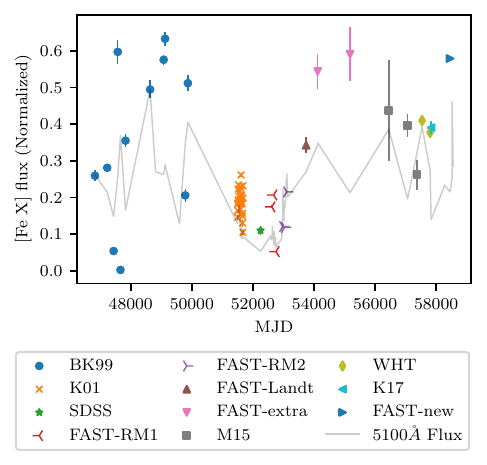}
    \caption{Fitted line fluxes for the \FeX{} emission line, with the error from fitting. The F5100 flux measured in \citetalias{homan2023} is shown for comparison. All fluxes are normalized against the narrow [O \textsc{i}] 6300 line flux.}
    \label{fig:fe10_time_series}
\end{figure}

Similar to the \FeVII{} line, it is possible to trace the evolution of the \FeX{} line flux, as shown in Figure \ref{fig:fe10_time_series}. However,  the fitting procedure of the \FeX{} line introduces difficult to quantify the errors associated with the presence of other features on the red wing of \FeX{}, our reported \FeX{} flux must be treated with some caution. Nonetheless, it does appear that the \FeX{} flux also declines, with a very short or zero time lag between it and the F5100 continuum. The \FeX{} flux in the K01 series is shown in \ref{fig:K01_time_series} and unlike \FeVII{}, it appears that \FeX{} does change over the duration of the K01 observations.

\subsection{Variability of the \NeV{} Line}

\begin{figure}
	\includegraphics[width=\columnwidth]{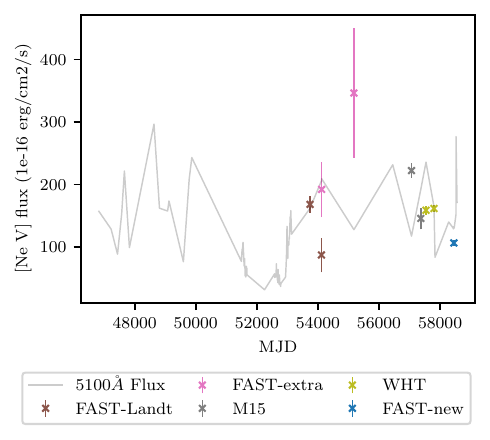}
    \caption{Fitted line fluxes for the \NeV{} emission line, along with the error from fitting. The F5100 flux measured in \citetalias{homan2023} is shown for comparison. The fitting error does not fully represent the total variance in the data. Each separate observing run is colour coded and labelled.}
    \label{fig:ne5_time_series}
\end{figure}

For completeness, where the spectrum includes the \NeV{} emission line, we have also fitted and measured the fluxes. Due to its wavelength in the blue optical, detections are limited to our later spectra, and so only some fluxes measured from the FAST-Landt spectra onwards are available (from January 2006). Despite this reduced time frame we do see variability in the \NeV{} flux, shown in Figure \ref{fig:ne5_time_series}.

\section{Quantifying the Lag with the F5100 Continuum} \label{sec:lag}
\subsection{The Lag of \FeVII{} Against the F5100 Continuum}

Inspection of Figure \ref{fig:time_series} suggests a decrease in \FeVII{} flux between about MJD 52000 and MJD 54000. This appears to be an "echo" of the dip in F5100 flux between MJD 51000 and MJD 53000. However, focusing on the K01 data (which has very good temporal sampling) as shown in Figure \ref{fig:K01_time_series}, we do not find a short term response of \FeVII{} to F5100. This suggests that the \FeVII{} flux is averaged out over this time period. We shall discuss this in more detail in Section \ref{sec:geometry}.

We estimated the lag between the \FeVII{} flux and the F5100 flux using two methods. The first is the Flux Randomization/Random Subset Selection (FR/RSS) method outlined in \citet{peterson1998}. A random number of epochs are selected (hence `Random Subset Selection') and their fluxes randomly perturbed by an amount which is normally distributed around the measured flux with a standard deviation equal to the fitting error (hence `Flux Randomization'). The cross-correlation curve between the sparse light curves is computed, and the centroid of the cross-correlation peak is reported as a lag. Repeating this repeatedly builds a distribution of lags. This enables not only an estimate of the lag, but also an error estimate.

\begin{figure}
	\includegraphics[width=\columnwidth]{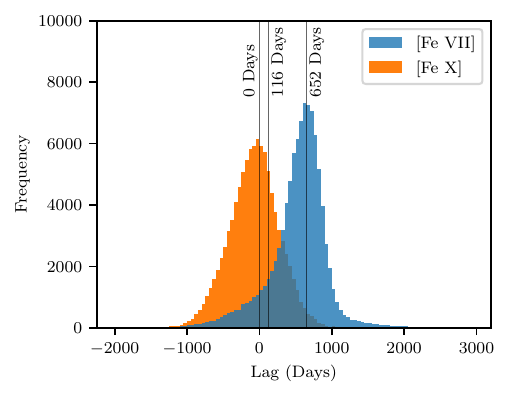}
    \caption{Distribution of Cross-Correlation Centroids using the FR/RSS method outlined in \citet{peterson1998} for both \FeVII{} and \FeX{}. 100,000 light curves were sampled for both \FeVII{} and \FeX{}. The highlighted lags are 652 days, the \FeVII{} lag result; 116 days, the lag of \FeX{} calculated using the \FeVII{} lag as well as the \FeVII{} and \FeX{} line widths; and 0 days.}
    \label{fig:fr_rss}
\end{figure}

Repeating the FR/RSS process 100,000 times, we show the full results in Figure \ref{fig:fr_rss}. There was a small secondary peak at $\sim$5000 days which we omitted on the grounds that since our data set spans about 10,000 days, we are unable to confidently consider lags on such a long time scale. 

\begin{figure}
	\includegraphics[width=\columnwidth]{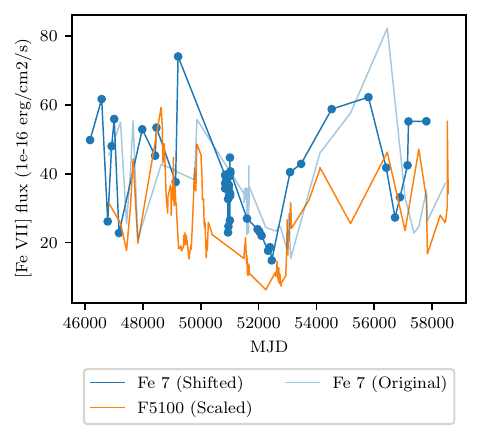}
    \caption{The \FeVII{} light curve shifted back by the proposed lag compared with the F5100 flux, which has been arbitrarily scaled for comparison purposes. The unshifted \FeVII{} light curve is also shown in lighter colour for comparison.}
    \label{fig:fe7_delagged}
\end{figure}

Omitting the secondary peak at $\sim$5000 days, we calculate a modal lag of 652 days. The median lag and 1-$\sigma$ errors (as derived from the lag distribution of sampled/simulated light curves) are $533_{-133}^{+113}$ days. Our confidence in this result is moderately high, as 90.5\% of the distribution corresponds to lags of greater than 0 days. Re-plotting the \FeVII{} light curve (as shown in Figure \ref{fig:fe7_delagged}) but shifting the F5100 flux back by the 652 days, the dip in F5100 and \FeVII{} flux appear much more closely aligned.

The second method used is JAVELIN \citep{zu2011}. This treats the continuum variation as a damped random walk, and assumes a top-hat transfer function between the continuum and observed emission line fluxes. JAVELIN then uses the Monte-Carlo Markov Chain approach to explore the parameter space of the damped random walk and of the transfer function allowing for lag estimation. Based on of the length of our light curve, we constrain the lag to be between 0 and 2000 days. With a total chain length of 100,000, the modal lag is 869 days. The median lag and 1-$\sigma$ errors are $871_{-6}^{+10}$ days. Given that only one major dip was covered in the series of observations, the continuum does not follow that of a damped random walk, and so the JAVELIN results should be treated with some caution. Nonetheless, it is noteworthy that a lag of the same order of magnitude (though not consistent) was produced.

\subsection{The Lag of \FeX{} Against the F5100 Continuum}

Using the FR/RSS method as for \FeVII{}, we can calculate a lag for \FeX{}. With the FR/RSS method, we calculate a modal lag of -41 days. The median lag and 1-$\sigma$ errors are $-65_{-147}^{+138}$, i.e. consistent with zero. The distributions of lags for both \FeVII{} and \FeX{} are shown in Figure \ref{fig:fr_rss}. The distributions are inconsistent with each other at 97.8\% confidence.

It is possible to arrive at an alternative figure for the \FeX{} lag using the \FeVII{} result along with the line widths discussed in \ref{sec:width}. The mean \FeVII{} velocity was \SI{330}{\kilo\meter\per\second} and the mean \FeX{} velocity was \SI{778}{\kilo\meter\per\second}. Scaling the \FeVII{} lag of 652 days using the relationship for virialized gas $v^2\sim r^{-1}$ gives a predicted lag of 116 days. This alternative lag is also highlighted in Figure \ref{fig:fr_rss}.

\section{The Responsivity of \FeVII{} against F5100}  \label{sec:responsivity}

\begin{figure}
	\includegraphics[width=\columnwidth]{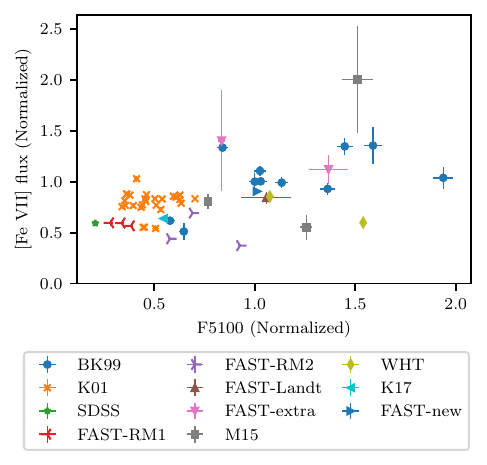}
    \caption{The fitted \FeVII{} fluxes versus the F5100 flux. All values are normalized against their values at MJD 47574. No corrections to take into account a lag are included.}
    \label{fig:scatter_nolag}
\end{figure}

\begin{figure}
	\includegraphics[width=\columnwidth]{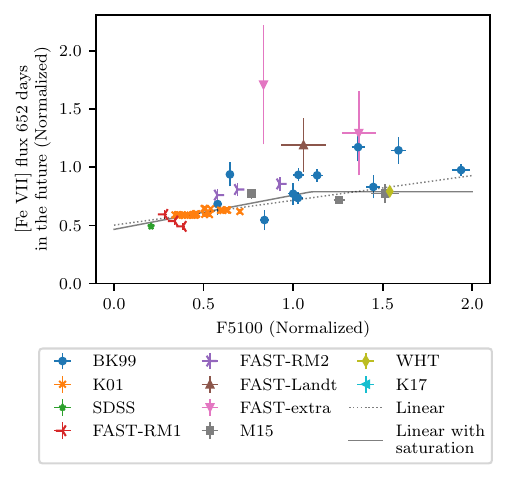}
    \caption{The fitted \FeVII{} flux corrected for lag versus the F5100 flux. For each date with \FeVII{} flux data, the F5100 flux 704 days ago was taken from the interpolated light curve. The 704 days corresponds to the modal FR/RSS lag. All values are normalized against their values at MJD 47574. A linear response with and without saturation is fitted and shown. The vertical outliers are likely a result of lower quality spectra - partially represented by the associated large error bars.}
    \label{fig:scatter_fe_704}
\end{figure}

To quantify the responsivity of \FeVII{} against the continuum, we plot the measured \FeVII{} fluxes against the corresponding F5100 flux. If we take "corresponding" to refer to the F5100 measured at the same time as the \FeVII{} flux, the results are shown in Figure \ref{fig:scatter_nolag}. Instead if we use the F5100 flux shifted by 652 days (the modal lag estimated using FR/RSS), the results are shown in Figure \ref{fig:scatter_fe_704}. In this manner the line fluxes are matches up with the continuum the line forming regions was exposed to at the time the lines were formed.

Accounting for a lag does result in a closer correlation between \FeVII{} and F5100. This is especially noticeable when examining the points corresponding to \FeVII{} from the SDSS, FAST-RM1 and FAST-RM2 spectra, which have very small errors.

To quantify the relationship between F5100 and \FeVII{}, we adopted a model similar to the saturated linear response in \citetalias{homan2023}, but with a non-varying component as a free parameter. This has the form:

\begin{equation}
	y(x) =
	\begin{cases}
		kx + c &                           \text{ if }x<x_\text{sat}\\
		y_\text{sat} = kx_\text{sat} + c & \text{ if }x\geq x_\text{sat}
	\end{cases}
\end{equation}

For comparison, we also fitted a simple linear model. Fitting both to the points in Figure \ref{fig:scatter_fe_704}, we tabulate the results in Table \ref{tab:fit}. 

\begin{table}
	\centering
	\caption{Fit parameters for the \FeVII{} - F5100 lag-adjusted responsivity.}
	\label{tab:fit}
	\begin{tabular}{ccc}
		\hline
		\multirow{2}{*}{Responsivity} & Non-responding & Saturating     \\
		                              & Component      & Ionizing Flux  \\
		$k$                           & $c$            & $x_\text{sat}$ \\
		\hline
		0.122 ± 0.123                 & 0.682 ± 0.067 & -               \\
		0.783 ± 0.092                 & 0.317 ± 0.051 & 0.840 ± 0.067   \\
		\hline
	\end{tabular}
\end{table}

The responses we found for both possible lag figures are less than 1. We find evidence of a non-varying component, represented by the offset $c$. This non-varying component appears to account for a substantial proportion of the total \FeVII{} flux.

There is some evidence for saturation of the \FeVII{} flux, such that beyond a certain point the \FeVII{} flux becomes unresponsive to changes in the F5100 flux. We fitted both a saturated model as well as a linear model, and found that accounting for the presence of saturation significantly improved the fit, reducing the $\chi^2$ statistic from 3607 to 1834. The large values of the $\chi^2$ statistics are likely caused by two main factors. The first is the lack of temporal smearing in this model, with the line emissions modelled as responding instantaneously to the continuum. The second is the presence of additional cosmic scatter that was not accounted for.

The variable fraction of the [FeVII] flux would likely be associated with a compact photionized zone. The more constant component could correspond to an extended region, possibly influenced by shocks.

\section{Discussion} \label{sec:discussion}

A Summary of the key results:
\begin{itemize}
	\item Both the \FeVII{} and \FeX{} fluxes vary, most likely in response to the \SI{5100}{\angstrom} continuum. However, they do not respond on short timescales in the way the BLR responds (on the order of days or weeks), but only on longer timescales (on the order of several months to years).
	\item We find clear evidence of a \FeVII{} lag of \SI{652}{\day}
	\item The \FeX{} lag is consistent with zero, and statistically much shorter than the \FeVII{} lag with a confidence of 89\%.
	\item The \FeX{} velocity width is larger than the velocity with of \FeVII{} by on average a factor of 2.
	\item The difference in lag and velocity widths both provide evidence of stratification of the coronal line region, as we discuss in detail in the following section. 
\end{itemize}

\subsection{Size of the Coronal Line Emitting Region - Evidence of Stratification}

There are two approaches to estimating the size of the \FeVII{} and \FeX{} emitting region, using either the lag or the velocity width, assuming that the observed width is a result of orbiting motion of virialized gas. Using both approaches, there is clear evidence that the \FeVII{} originates predominantly from a different region than the \FeX{} region.

Using the lag results, we estimate a \FeVII{} emitting region size of 652 light days. As distribution for \FeVII{} lags and \FeX{} lags are clearly inconsistent - 97.8\% of the \FeX{} lag distribution lies below the 652 days for \FeVII{} lag, the size of the \FeX{} emitting region must be smaller than that of the \FeVII{{} emitting region.

Using line widths and the classic reverberation mapping relationship requires a value for the black hole mass. However, there are two possible but distinct black hole masses reported for MKN 110 - \SI{1.8e7}{\msol} and \SI{1.4e8}{\msol}. The first was derived from the broad line widths and the lag. The second was derived by assuming that broad line velocity offsets are caused by a gravitational redshift. \citet{kollatschny2003} reconciles this difference by suggesting an inclination angle of \SI{21}{\degree} as measured from pole-on. For \FeVII{} with a typical velocity width of \SI{370}{\kilo\meter\per\second}, this corresponds to a physical size of 652 light days. For \FeX{} with a typical velocity width of \SI{700}{\kilo\meter\per\second}, this corresponds to a physical size of 116 light days.

The \FeVII{} results for physical size reported here, with the exception of those derived from using JAVELIN, are all consistent. The failure of JAVELIN results to agree may be a result of the caveats discussed earlier regarding using that technique. The \FeX{} results however are not consistent with the FR/RSS lag, but taking into account the large uncertainties of the \FeX{} measurements, this may be a result of the relatively poor quality of the available data.

Comparing our results with previous studies of coronal lines, we note in general there are two locations which have been cited as possible emitting regions. Spatially resolved imaging by, for instance, \citet{rodriguez-ardila2006} and \citet{mazzalay2010} suggests the CLR is at distances of up to $\sim\SI{100}{\parsec}$, co-spatial with the resolved narrow line region. Indeed \citet{rodriguez-ardila2020} has found coronal line emission \SI{700}{\parsec} from the central object in the Circinus galaxy. A portion of \FeVII{} emission in the Circinus galaxy appears to originate in clumps, likely outflows. The other suggestion, by \citet{mullaney2009}, \citet{rose2015} and \citet{adhikari2016}, is a location of about $\sim\SI{1}{\parsec}$, between the BLR and NLR, and possibly associated with the inner wall of the dusty "torus". This is supported by studies of velocity profiles together with photoionization modelling. We note, as \citet{murayama1998} suggested, that these locations for \FeVII{} are not contradictory, and are likely to co-exist under different physical conditions. The survey of coronal line emitting AGNs by \citet{negus2023} identifies objects with \FeVII emitting regions from both the nucleus and extended clumps, and in some rare cases only the extended region. While MRK110 is not included in the sample by \citet{negus2023}, it is perhaps not unreasonable to suggest that MRK110 is similar.

\subsection{The Physical Condition of the CLR} \label{sec:modelling}

Based on very general assumptions it is not surprising that coronal lines are emitted over a large range of distances from the central source. It is now well established from previous modelling \citep[for example by][]{ferguson1997} that the key dimensionless parameter in emission line reprocessing is the "ionization parameter" $U$, given by:
\begin{equation}
	U = \frac{Q}{4\pi r^2 n c}
\end{equation}
where $Q$ is the number of ionizing photons produced by the central source, $r$ the radius of the cloud of material, $n$ the hydrogen density and $c$ the speed of light. In principle for a given coronal line at any distance, it is possible for a cloud with the required particle density to produce that line.

In practice, based on our lag result, we can constrain the distance of \FeVII{} and \FeX{} emitting clouds. This would then allow us to estimate the density of \FeVII{} and \FeX{} emitting clouds.

To demonstrate the feasibility of \FeVII{} and \FeX{} emission at various densities, we present a simple model of gas in the CLR using \cloudy{} \citep{chatzikos2023}. For convenience, we used the default \cloudy{} AGN SED, normalized to match the observed continuum of MKN 110. The density was varied from between \SI{1e2}{\per\centi\meter\cubed} to \SI{1e9}{\per\centi\meter\cubed}. This range corresponds to densities generally accepted for the NLR and the BLR. The illuminated side of the \FeVII{} cloud was positioned at a distance of \SI{2e18}{\centi\meter} or 643 light days from the central source. The illuminated side of the \FeX{} cloud was positioned at a distance of \SI{3e17}{\centi\meter} or 116 light days from the central source. This distance comes from taking the lag distance of 652 light days, and reducing it based on measurements of the profile velocity widths of \FeVII{} and \FeX{}.

Two options are available for setting the outermost/furthest boundary of the cloud - either an arbitrary proportion of the distance to the ionizing source or when the temperature at the outer boundary drops below \SI{4000}{\kelvin}, whereupon no further optical emission is expected. The first is often described as "Matter Limited", where the factor limiting line flux is the amount of matter to ionize. The second is often described as "Radiation Limited", where the line flux is limited by the amount of ionizing radiation. Both possibilities are shown in Figure \ref{fig:agn_flux}. For the purposes of this model, we chose the boundary of the cloud to be 0.1 the distance to the ionizing source for the "Matter Limited" case, yielding clouds uniformly \SI{2e17}{\centi\meter} thick for \FeVII{} and \SI{3e16}{\centi\meter} for \FeX{}. Changing this factor only affects the density where the matter and radiation limited models diverge.

\begin{figure}
	\includegraphics[width=\columnwidth]{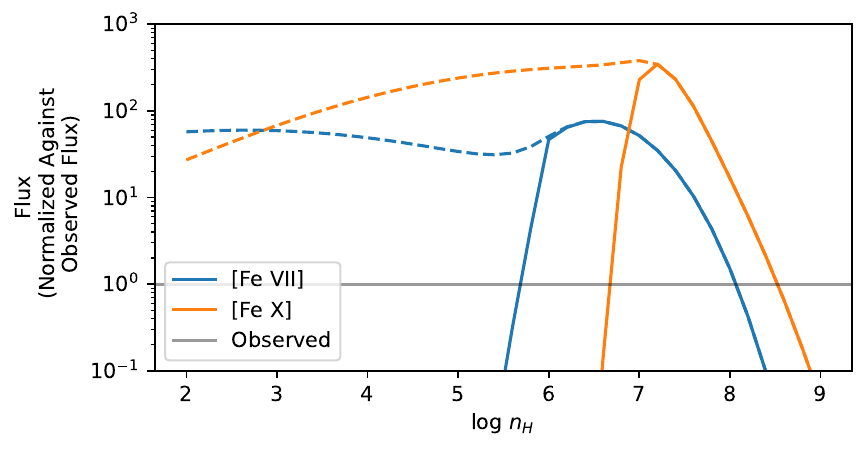}
    \caption{\FeVII{} and \FeX{} fluxes as predicted by \cloudy{}, normalized against the respective observed fluxes, for both  matter and radiation limited models. The dashed lines represent the radiation limited \cloudy{} models, and the solid lines represent the matter limited \cloudy{} models. See text for further details. Values greater than $1$ represent densities where \cloudy{} was able to reproduce the observed flux.}
    \label{fig:agn_flux}
\end{figure}

For both \FeVII{} and \FeX{} radiation limited \cloudy{} models can produce sufficient line emission across a wide range of densities, but the matter limited \cloudy{} models were only able to produce sufficient line emission over a narrower range of densities. The \FeVII{} and \FeX{} flux in matter limited \cloudy{} models show a rapid drop off below a certain density. The behaviour of radiation limited and matter limited \cloudy{} models converge to be the same at high densities.

However, since the only stopping condition in the radiation limited \cloudy{} models is the gas temperature, it produces clouds with back edges $\sim\SI{1e22}{\centi\meter}$ or further from the central source. In these clouds, the Fe+6 region (which emits \FeVII{}) in lower density clouds, can be found at $\sim\SI{1e20}{\centi\meter}$ corresponding to $\sim\SI{10}{\parsec}$ from the illuminated face of the cloud. Clearly, a \FeVII{} emitting region this far out is completely inconsistent with the lag results that we have found. Furthermore, as \citet{mckaig2024} points out, for gas in the dust the region which exists beyond the sublimation radius, the coronal line emission is weaker. This needs to be considered further when applying lower density radiation-limited models.

The sharp drop in \FeVII{} and \FeX{} flux below a certain density in the "Matter Limited" case is caused by the cloud being over ionized, with Fe+7, +8 etc. or Fe+10, +11 etc. being produced instead of Fe+6 and Fe+9 respectively. Although it is possible to adjust the thickness of the clouds in order to produce more \FeVII{} and \FeX{} flux, as discussed above, lag results provide a constraint on the maximum thickness possible. In the case of this simple model, the \FeVII{} emitting cloud located at a distance consistent with the lag result must have a density between $n_H \sim \SI{5e5}{\per\centi\meter\cubed}$ and $n_H \sim \SI{1e8}{\per\centi\meter\cubed}$. The \FeX{} emitting cloud must have a density between $n_H \sim \SI{5e6}{\per\centi\meter\cubed}$ and $n_H \sim \SI{5e8}{\per\centi\meter\cubed}$. This difference in emitting densities remains even if the clouds are located at the same distance. It should be noted that these results are illustrative only, and are not proposed as actual predictions of \FeVII{} and \FeX{} fluxes. An full attempt at modelling the CLR in MRK110 would require a more accurate model of the SED, as well as other effects such radiation pressure compression, and dust grains as mentioned above.

Nonetheless, the fact that these lower density clouds can produce \FeVII{} and \FeX{} emissions, albeit much further out, may explain some of the spatially resolved coronal line emission at the \qtyrange{10}{100}{\parsec} scale. Since lower density clouds further out in the NLR are capable of producing coronal line emission, this region may be the source of the non-varying component described in Section \ref{sec:responsivity}. Clumpy clouds of higher density within an environment of lower density gas may provide the innermost coronal line emission, with lines from more highly ionized species produced closer to the centre of the AGN. Further out clouds of lower density would also produce coronal lines, and at distances of the NLR the coronal lines mostly emanate from the gas between narrow line emitting clouds.

In summary, the results of our modelling enable us to build a consistent picture of a stratified CLR. The \FeX{} emitting region should have higher densities than the \FeVII{} emitting region, which suggests a radial density profile. The simplest way to represent the radial dependence of the cloud density is with a power law:
\begin{equation}
	n_H \sim r^{-\beta}
\end{equation}
where $\beta$ is a free parameter. So if $\beta = 2$ within the CLR, with ionization parameter $U = L/4\pi r^2 nc$, $U$ has no radial dependence. But this is not consistent with our results for lag and a stratified CLR. If instead we take densities resulting in the minimum implied covering factor (i.e. the maximum flux) for both \FeVII{} and \FeX{}, we find that there is a difference of 0.6 dex difference in $\log n_H$ between the \FeVII{} and \FeX{} emitting clouds. Taking 643 light days and 116 light days as the distances for \FeVII{} and \FeX{} imply $\beta = 0.81$. We also consider where the implied covering factor crosses $c_F = 1$ (i.e. the flux predicted by \cloudy{} matches the observed flux). With the two sets of solutions, we obtain either $\beta = 1.33$ or $\beta = 0.62$.

\subsection{Geometry of the CLR} \label{sec:geometry}

Velocity resolved BLR reverberation mapping is capable of mapping the velocity fields of the BLR of AGNs. However, in the case of MKN110 given the limited spectral resolution and time sampling resolution, we are unable to perform such analysis using the spectra available to us. Theoretical calculations of transfer functions such as those by \citet{perez1992} have found that all commonly suggested geometries (spheres, disk or bi-polar cones) produces 1D transfer functions with centroids equal to the luminosity weighted radius. Therefore velocity resolved emission line reverberation mapping is required in order to discriminate between different geometries.

Nonetheless, it is still possible to draw some general conclusions. We observe a \FeVII{} responsivity $< 1$, plus the absence of a short-term response (demonstrated in Figure \ref{fig:K01_time_series}). Both these results suggest a broad transfer function, which in turn indicates that the \FeVII{} emitting region is either spatially diffuse or distributed over a large range of angles when viewed from the central source, and probably both simultaneously. Furthermore the responsivity results reveal the presence of a non varying component, which may also be explained by \FeVII{} emission produced much further out than the FR/RSS lag. This is supported by observations of a number of AGN \citep{mazzalay2010}. They find most objects have coronal line emission concentrated within the innermost $\sim\SI{10}{\parsec}$, limited by spatial the resolution, but often with coronal line emission extended much further out into the NLR.

\section{Conclusions}

MKN 110 is one of the few objects with available spectra spanning over several decades. This long time series of spectra combined with a broad and sustained dip in continuum flux allowed us to demonstrate that the \FeVII{} coronal line clearly varies, and responds to the continuum. The \FeVII{} variability and response occurs over a long time scale of years, but does not occur on shorter daily or weekly timescales. This is in contrast to the variability and response of the broad emission lines and the continuum, as discussed for MKN110 in \citetalias{homan2023}.

We find the \FeVII{} flux lags the F5100 continuum by 652 days. In terms of responsivity, there is evidence of a non-varying component, as well as saturation, where increasing continuum flux does not result in corresponding increases in \FeVII{} flux. Where the \FeVII{} flux does respond to the continuum, the slope of the response is $<1$.

The \FeVII{} velocity widths are only just separable from the instrumental resolution, with a mean of \SI{330}{\kilo\meter\per\second}. The mean \FeX{} width is broader with \SI{778}{\kilo\meter\per\second}. While the \FeX{} flux does vary, we did not detect any significant lag for this line. The distribution of possible lags of \FeVII{} and \FeX{} are inconsistent with each other. The width of \FeX{}, together with the width and lag of \FeVII{} can be used to estimate an \FeX{} distance of 116 days. However this value is not inconsistent with the lag distance distribution for \FeX{} calculated using the FR/RSS method.

The observed lag, the line widths, and our basic photoionisation modelling indicate the presence of a stratified CLR, with the \FeX{} produced closer to the central source than \FeVII{}. This is in agreement with previous studies of samples of AGN with coronal lines such as by \citet{gelbord2009} and \citet{rodriguez-ardila2004}. We conclude that when attempting to use observations of coronal lines to constrain shape of the SED in the EUV, a simple single-cloud model is insufficient to explain the properties of a stratified region extending from the outermost BLR into the classic NLR. In future more sophisticated modelling is needed, to include velocity width flux dependence and spatial density profile.

\section*{Acknowledgements}

WK greatly acknowledges support by the DFG grant KO 857/35-1.

Some spectra are based on observations obtained with the Hobby-Eberly Telescope, which is a joint project of the University of Texas at Austin, the Pennsylvania State University, Ludwig-Maximilians-Universit\"at M\"unchen, and Georg-August-Universit\"at G\"ottingen.

\section*{Data Availability}

Any additional line flux data available upon request.

For the purpose of open access, the author has applied a Creative Commons Attribution (CC BY) licence to any Author Accepted Manuscript version arising from this submission.




\bibliographystyle{mnras}
\bibliography{bib}




\appendix

\section{Fitting Results for \FeVII{}}

\begin{table}
	\centering
	\caption{Our line fit parameters for \FeVII{}.}
	\label{tab:fe7_results}
	\begin{tabular}{cccc}
		\hline
		MJD   & Flux & Central Wavelength & Width    \\
		\hline
		46828 & 45.5 & 6086.7             & 5.7      \\
		47229 & 54.9 & 6088.0             & 7.0      \\
		47438 & 25.4 & 6085.4             & 5.4      \\
		47574 & 41.1 & 6087.2             & 4.7      \\
		47663 & 55.4 & 6089.9             & 6.5      \\
		47828 & 21.1 & 6087.6             & 4.3      \\
		48632 & 42.7 & 6087.9             & 8.0      \\
		49078 & 41.3 & 6088.1             & 4.2      \\
		49123 & 40.8 & 6084.4             & 3.7      \\
		49785 & 38.3 & 6086.8             & 2.6      \\
		49870 & 55.7 & 6087.6             & 5.9      \\
		51495 & 34.3 & 6087.2             & 4.6      \\
		51497 & 31.6 & 6087.0             & 4.6      \\
		51500 & 34.2 & 6087.1             & 4.9      \\
		51518 & 33.9 & 6087.1             & 5.4      \\
		51520 & 35.8 & 6087.2             & 5.1      \\
		51522 & 32.5 & 6086.9             & 4.9      \\
		51525 & 35.3 & 6087.1             & 4.8      \\
		51528 & 34.9 & 6087.6             & 4.6      \\
		51547 & 34.3 & 6087.3             & 5.1      \\
		51584 & 35.9 & 6086.1             & 5.0      \\
		51586 & 34.2 & 6087.0             & 4.7      \\
		51595 & 22.3 & 6088.2             & 4.8      \\
		51598 & 29.9 & 6086.8             & 5.7      \\
		51605 & 33.2 & 6087.1             & 5.5      \\
		51608 & 31.8 & 6086.4             & 5.2      \\
		51611 & 22.7 & 6088.3             & 4.8      \\
		51614 & 31.6 & 6086.9             & 4.8      \\
		51629 & 33.5 & 6087.6             & 4.8      \\
		51637 & 31.1 & 6086.8             & 4.8      \\
		51645 & 35.7 & 6086.9             & 5.0      \\
		51658 & 22.8 & 6087.1             & 3.9      \\
		51663 & 42.3 & 6086.2             & 4.9      \\
		51664 & 31.5 & 6088.5             & 4.5      \\
		51670 & 30.8 & 6086.9             & 3.9      \\
		51673 & 36.1 & 6086.3             & 5.2      \\
		51678 & 36.3 & 6087.2             & 4.8      \\
		52252 & 24.4 & 6088.2             & 3.7      \\
		52617 & 23.3 & 6085.3             & 3.2      \\
		52679 & 24.5 & 6085.2             & 3.1      \\
		52755 & 24.5 & 6084.1             & 3.6      \\
		52979 & 18.1 & 6085.1             & 3.8      \\
		53043 & 28.5 & 6084.8             & 3.8      \\
		53108 & 15.3 & 6085.1             & 3.7      \\
		53741 & 34.7 & 6086.8             & 3.9      \\
		54122 & 46.0 & 6088.7             & 2.9      \\
		55179 & 57.6 & 6084.5             & 8.0      \\
		56450 & 82.3 & 6087.3             & 2.5      \\
		57064 & 33.2 & 6087.3             & 3.4      \\
		57369 & 22.8 & 6088.6             & 4.4      \\
		57539 & 24.6 & 6087.8             & 3.0      \\
		57801 & 35.2 & 6086.8             & 4.6      \\
		57834 & 26.3 & 6087.6             & 4.1      \\
		58452 & 37.2 & 6086.1             & 4.0      \\
				\hline
	\end{tabular}
\end{table}

\section{Fitting Results for \FeX{}}

\begin{table}
	\centering
	\caption{Our line fit parameters for \FeX{}.}
	\label{tab:fe10_results}
	\begin{tabular}{cccc}
		\hline
		MJD   & Flux  & Central Wavelength & Width    \\
		\hline
		46828 &  47.3 & 6380.0             & 9.0      \\
		47229 &  49.9 & 6377.4             & 9.0      \\
		47438 &  10.4 & 6380.0             & 5.6      \\
		47574 & 102.4 & 6377.1             & 9.0      \\
		47663 &   0.4 & 6377.1             & 1.0      \\
		47828 &  65.6 & 6371.9             & 6.1      \\
		48632 &  79.8 & 6374.9             & 9.0      \\
		49078 & 105.0 & 6376.4             & 9.0      \\
		49123 &  96.6 & 6371.4             & 9.0      \\
		49785 &  41.8 & 6375.1             & 9.0      \\
		49870 &  77.0 & 6375.1             & 9.0      \\
		51495 &  25.3 & 6375.8             & 6.9      \\
		51497 &  32.2 & 6376.0             & 9.0      \\
		51500 &  30.1 & 6375.1             & 7.1      \\
		51518 &  33.6 & 6374.7             & 8.2      \\
		51520 &  40.0 & 6374.6             & 9.0      \\
		51522 &  35.1 & 6374.9             & 8.6      \\
		51525 &  35.4 & 6374.9             & 8.6      \\
		51528 &  43.5 & 6374.7             & 9.0      \\
		51547 &  40.3 & 6375.6             & 9.0      \\
		51584 &  35.6 & 6375.3             & 7.5      \\
		51586 &  37.0 & 6375.6             & 8.1      \\
		51595 &  32.5 & 6378.8             & 9.0      \\
		51598 &  41.2 & 6374.6             & 9.0      \\
		51605 &  38.0 & 6375.3             & 8.7      \\
		51608 &  35.2 & 6374.6             & 6.9      \\
		51611 &  48.0 & 6375.0             & 9.0      \\
		51614 &  35.7 & 6375.4             & 9.0      \\
		51629 &  37.6 & 6375.2             & 9.0      \\
		51637 &  34.4 & 6375.7             & 9.0      \\
		51645 &  33.1 & 6375.2             & 7.4      \\
		51658 &  26.9 & 6372.9             & 6.8      \\
		51663 &  24.7 & 6375.7             & 7.8      \\
		51664 &  27.4 & 6376.5             & 8.4      \\
		51670 &  42.4 & 6374.2             & 9.0      \\
		51673 &  18.9 & 6375.6             & 5.6      \\
		51678 &  35.3 & 6373.4             & 9.0      \\
		52252 &  19.9 & 6375.0             & 5.6      \\
		52617 &  34.0 & 6371.1             & 9.0      \\
		52679 &  43.5 & 6375.1             & 8.3      \\
		52755 &  11.6 & 6374.9             & 2.3      \\
		52979 &  24.8 & 6373.2             & 5.3      \\
		53043 &  36.0 & 6373.0             & 4.8      \\
		53108 &  44.4 & 6374.6             & 9.0      \\
		53741 &  58.6 & 6374.5             & 8.4      \\
		54122 & 116.7 & 6374.0             & 8.5      \\
		55179 & 115.9 & 6372.1             & 9.0      \\
		56450 & 115.7 & 6370.8             & 9.0      \\
		57064 &  63.0 & 6371.5             & 8.1      \\
		57369 &  43.7 & 6373.0             & 4.7      \\
		57539 &  60.8 & 6375.5             & 9.0      \\
		57801 &  62.5 & 6375.8             & 9.0      \\
		57834 &  37.2 & 6374.2             & 7.4      \\
		58452 &  78.0 & 6371.5             & 9.0      \\
		\hline
	\end{tabular}
\end{table}

\section{Fitting Results for [O \textsc{i}] \SI{6300}{\angstrom}}

\begin{table}
	\centering
	\caption{Our line fit parameters for [O \textsc{i}] $\lambda$\SI{6300}{\angstrom}.}
	\label{tab:o6300_results}
	\begin{tabular}{cccc}
		\hline
		MJD & Flux & Central Wavelength & Width    \\
		\hline
		46828 & 182.7 & 6301.2           & 6.0     \\
		47229 & 178.0 & 6301.2           & 5.5     \\
		47438 & 194.3 & 6300.0           & 6.1     \\
		47574 & 171.4 & 6301.2           & 4.5     \\
		47663 & 198.2 & 6300.5           & 5.1     \\
		47828 & 184.8 & 6300.4           & 4.2     \\
		48632 & 161.4 & 6301.1           & 8.2     \\
		49078 & 182.4 & 6301.2           & 4.5     \\
		49123 & 152.7 & 6297.3           & 3.9     \\
		49785 & 203.5 & 6300.1           & 3.8     \\
		49870 & 150.5 & 6301.5           & 4.2     \\
		51495 & 173.4 & 6301.2           & 4.4     \\
		51497 & 177.1 & 6301.1           & 4.6     \\
		51500 & 183.6 & 6301.0           & 4.4     \\
		51518 & 183.0 & 6300.9           & 4.5     \\
		51520 & 179.8 & 6301.0           & 4.4     \\
		51522 & 181.4 & 6301.0           & 4.4     \\
		51525 & 187.2 & 6301.0           & 4.4     \\
		51528 & 185.2 & 6300.9           & 4.4     \\
		51547 & 186.4 & 6300.7           & 4.4     \\
		51584 & 184.1 & 6301.2           & 4.5     \\
		51586 & 183.8 & 6301.0           & 4.4     \\
		51595 & 193.9 & 6300.7           & 4.6     \\
		51598 & 183.0 & 6301.1           & 4.4     \\
		51605 & 183.4 & 6301.4           & 4.3     \\
		51608 & 179.9 & 6300.7           & 4.2     \\
		51611 & 183.7 & 6301.1           & 4.5     \\
		51614 & 178.8 & 6301.1           & 4.4     \\
		51629 & 182.8 & 6301.2           & 4.3     \\
		51637 & 183.5 & 6301.0           & 4.5     \\
		51645 & 181.6 & 6301.6           & 4.3     \\
		51658 & 172.1 & 6300.9           & 4.3     \\
		51663 & 189.4 & 6300.4           & 4.6     \\
		51664 & 183.5 & 6302.7           & 4.4     \\
		51670 & 183.2 & 6300.9           & 4.2     \\
		51673 & 180.5 & 6301.0           & 4.3     \\
		51678 & 178.2 & 6301.0           & 4.3     \\
		52252 & 180.8 & 6302.2           & 3.2     \\
		52617 & 195.3 & 6299.8           & 3.6     \\
		52679 & 210.5 & 6300.5           & 3.7     \\
		52755 & 223.0 & 6301.1           & 3.6     \\
		52979 & 205.3 & 6300.3           & 3.6     \\
		53043 & 305.0 & 6300.4           & 3.6     \\
		53108 & 206.5 & 6300.8           & 3.6     \\
		53741 & 171.3 & 6300.8           & 3.5     \\
		54122 & 214.7 & 6300.7           & 3.4     \\
		55179 & 196.1 & 6299.7           & 2.9     \\
		56450 & 264.5 & 6302.2           & 4.2     \\
		57064 & 158.8 & 6300.8           & 3.4     \\
		57369 & 166.9 & 6302.1           & 4.3     \\
		57539 & 148.4 & 6301.1           & 3.8     \\
		57801 & 165.7 & 6301.0           & 4.6     \\
		57834 & 95.2  & 6301.1           & 3.2     \\
		58452 & 134.8 & 6299.6           & 3.3     \\
		\hline
	\end{tabular}
\end{table}


\bsp	
\label{lastpage}
\end{document}